\documentclass[aps,twocolumn,groupedaddress,showpacs,showkeys]{revtex4}

\newcommand{\tr}{{\textrm {tr}}}

\newcommand{\Det}{{\textrm {Det}}}
\newcommand{\D}{{\textrm {D}}}
\newcommand{\A}{{\mathcal A}}

\newcommand{\bfp}{\mbox{\boldmath $p$}}
\newcommand{\bfx}{\mbox{\boldmath $x$}}

\begin{document}

\title{The induced Chern-Simons term at finite temperature}

\author{L.L. Salcedo}
\email{salcedo@ugr.es}

\affiliation{
Departamento de F\'{\i}sica Moderna,
Universidad de Granada,
E-18071 Granada, Spain
}

\date{\today} 

\begin{abstract}
It is argued that the derivative expansion is a suitable method to
deal with finite temperature field theory, if it is restricted to
spatial derivatives only. Using this method, a simple and direct
calculation is presented for the radiatively induced
Chern-Simons--like piece of the effective action of (2+1)-dimensional
fermions at finite temperature coupled to external gauge fields. The
gauge fields are not assumed to be subjected to special constraints,
and in particular, they are not required to be stationary nor Abelian.
\end{abstract}


\maketitle

\section{Introduction}
Finite temperature field theory \cite{Landsman:1987uw} is notoriously
more difficult to deal with than its zero temperature
counterpart. Lorentz symmetry is reduced and time and space play
different roles. This is apparent in the imaginary time formalism, in
which the time is Wick-rotated and compactified to a circle. As a
consequence, tools such as the perturbative expansion or the
derivative expansion, that were quite useful at zero temperature may
become unreliable. These remarks apply, in particular, to
$(2+1)$-dimensional QED. In this theory, due to the compactification,
topologically non-trivial (large) gauge transformations are supported
at finite temperature \cite{Pisarski:1987gq}. In principle, one would
expect that the coefficient of the radiatively induced Chern-Simons
(CS) term in the effective action \cite{Redlich:1984kn} would be
correctly quantized \cite{Deser:1982wh}, consistently with gauge
invariance of the partition function at any temperature
\cite{Deser:1997nv}. This turns out not to be the case when that
coefficient is computed using perturbation theory
\cite{Pisarski:1987gq,Babu:1987rs}. Existing (Abelian and non-Abelian)
exact results for particular configurations
\cite{Dunne:1997yb,Deser:1997nv,Fosco:1997ei,Brandt:2001ni} show that
gauge invariance is an exact symmetry but perturbation theory fails to
see this; perturbation theory is designed to describe the effective
action in the neighborhood of a vanishing gauge field and large gauge
transformations necessarily move the gauge field configuration away
from the perturbative region. The derivative expansion is also
problematic as already shown in a $(0+1)$-dimensional setting
\cite{Barcelos-Neto:1998xh} (using the real time formalism). The
problem is clear in the imaginary time formalism since the energy
becomes a discrete variable. This suggests that the trouble exists
only if one insists in taking a derivative expansion in both time and
space. Technically, to make a derivative expansion of the effective
action functional $W$, means to replace the given configuration
$A_\mu(x)$ by a family of configurations $A_\mu(\lambda x)$, where
$\lambda$ is a bookkeeping parameter, and then expand $W(\lambda)$ in
powers of $\lambda$ \footnote{This shows that each term of the
expansion is independent of the {\em method} used in the
calculation. This fails to hold for other expansions, such as the
commutator expansion.}. At finite temperature, this procedure is
inconsistent, since the external (bosonic) fields are required to be
periodic in the (Euclidean) time and the dilatation in the temporal
direction breaks this constraint. On the other hand, no problem should
appear if the derivative expansion is restricted to the spatial
directions (or at least, no new pathologies as compared to the zero
temperature case). Moreover, gauge invariance can be accounted for if
the bookkeeping parameter is introduced in the form
$A_0(x_0,\lambda\bfx)$, $\lambda A_i(x_0,\lambda\bfx)$.

The exact results in \cite{Deser:1997nv,Fosco:1997ei,Brandt:2001ni}
were obtained for particular configurations of the gauge field using
specially adapted methods. In this Letter we show that general
configurations are also amenable to explicit calculation using the
method of expanding the effective action in the number of spatial
covariant derivatives. We present a direct calculation of the induced
CS term at finite temperature which yields a simple explicit
expression for this quantity. This expression holds for general
configurations and agrees with all previously known exact results. Of
course, in addition to this, the effective action contains further
contributions which are completely regular (i.e. gauge and parity
invariant and ultraviolet finite) which are also calculable within the
same general scheme.

The fermionic effective action is defined as $W=-\log\Det{\D}$, where
$\D =\gamma_\mu D_\mu+m$ is the Dirac operator, $m$ is the fermion
mass and $D_\mu=\partial_\mu+A_\mu$ is the covariant derivative
\footnote{Our conventions follow those in
\cite{Salcedo:1998sv,Garcia-Recio:2000gt} with $\eta=+1$.}. The Dirac
operator acts on a single-particle Hilbert space which contains
space-time, Dirac and internal degrees of freedom (i.e. those
associated to the gauge group). The finite temperature $T=1/\beta$ is
introduced by compactifying the Euclidean time. As usual the fermionic
wave functions are antiperiodic and the gauge fields are periodic in
the time variable.

As is well known, the effective action is afflicted by several related
pathologies, namely, ultraviolet divergences, many-valuation and
anomalies in some of the classical symmetries of the Dirac
operator. Our purpose is to isolate precisely those pieces of the
effective action which can have an anomalous or many-valued
contribution. This excludes ultraviolet finite pieces, which are
always regular.  The expansion in the number of spatial covariant
derivatives \cite{Salcedo:1998sv,Garcia-Recio:2000gt} is appropriate
for this kind of calculations, since this expansion preserves gauge
invariance order by order and, in addition, terms beyond second order
are ultraviolet finite. A further simplification can be achieved by
selecting only those terms which have abnormal parity, i.e., those
containing a Levi-Civita pseudo-tensor, since the normal parity
component of the effective action can be renormalized preserving all
classical symmetries and is one-valued \footnote{This remark applies
to the massive case only. In the massless case \cite{Redlich:1984kn},
or if $|\mu|>|m|$ (where $\mu$ is the chemical potential)
\cite{Sisakian:1998cp}, the normal parity term combines with the
abnormal parity one to yield a one-valued partition functional
$\Det{\D}$.}. In passing we note that ``parity'' (defined as space
reflection but including $m\to -m$) is a symmetry of the classical
action. Therefore, we would expect the normal (abnormal) parity
component of the effective action to be an even (odd) function of the
mass. This symmetry may be anomalously broken as a consequence of
gauge invariance, as in the massless case \cite{Redlich:1984kn}.

\section{The current}
Following Schwinger \cite{Schwinger:1951nm}, we will work with the
current, which is better behaved than the effective action. The
current is defined through the relation $\delta W = \int d^3x\, \tr
(J_\mu \delta A_\mu)$ (where $\delta A_\mu$ is a local variation,
$J_\mu(x)$ is a matrix in internal space and $\tr$ refers to this
space), and thus
\begin{equation}
J_\mu(x)= -\tr_D\langle x| \gamma_\mu \frac{1}{\gamma_\nu
D_\nu+m}|x\rangle\,.
\end{equation}
($\tr_D$ refers to Dirac space.) In view of our previous remarks, our
approach will be to compute the relevant terms of the spatial current
$J_i$ and then reconstruct the effective action from it. This requires
to expand the current in powers of $D_i$ retaining only terms with
$\epsilon_{ij}$ and with precisely one $D_i$.  A suitable way to deal
with the matrix element at coincident points, which combines well with
gradient expansion, is to use the method of symbols
\cite{Salcedo:1996qy}, adapted to finite temperature
\cite{Salcedo:1998sv,Garcia-Recio:2000gt} and improved by Pletnev and
Banin \cite{Pletnev:1998yu,Salcedo:2000hx}. This gives
\begin{equation}
J_i(x)= -  \frac{1}{\beta}\sum_{p_0}\int\frac{d^2\bfp}{(2\pi)^2}
\,\tr_D\langle x|\gamma_i  \frac{1}{\gamma_\nu \tilde{D}_\nu+m}|0\rangle \,.
\label{eq:1}
\end{equation}
In this formula $|0\rangle$ is the constant wave function, i.e.
$\langle x| 0\rangle= 1$ (which is periodic rather than antiperiodic),
$p_0= 2\pi(n+\frac{1}{2})/\beta$ runs over the fermionic Matsubara
frequencies, and finally, $\tilde{D}_\mu$ is related to the original
covariant derivative through a double similarity transformation,
\begin{equation}
\tilde{D}_\alpha
= e^{i\partial^p_i D_i}e^{-ix_\mu p_\mu} D_\alpha
 e^{ix_\nu p_\nu} e^{-i\partial^p_j D_j}\,,\quad
(\partial^p_i= \frac{\partial}{\partial p_i}).
\label{eq:12}
\end{equation}
The inner similarity transformation yields $D_\mu\to D_\mu + ip_\mu$
and corresponds to the original method of symbols. In that method, the
integration over $\bfp$ cancels all contributions except those where
$D_i$ appears inside a commutator (more precisely, in the form $[D_i,\
]$). In the improved method, this
cancellation is achieved prior to momentum integration by means of the
outer similarity transformation in (\ref{eq:12}). Explicitly:
\begin{eqnarray}
\tilde{D}_0 &=& ip_0 + D_0 - i E_i \partial^p_i + \frac{1}{2}
[D_i,E_j] \partial^p_i \partial^p_j
+ O(D_i^3)\,, \nonumber \\
\tilde{D}_i &=& ip_i  +\frac{i}{2}F_{ij} \partial^p_j +  O(D_i^3) \,,
\label{eq:11}
\end{eqnarray}
where $F_{\mu\nu}=[D_\mu,D_\nu]$ and $E_i=F_{0i}$ is the electric
field. The improvement does not extend to the compactified time
coordinate, and indeed at finite temperature $D_0$ appears in two
different ways which preserve gauge invariance, i) inside commutators,
$[D_0,\ ]$, and ii) through the (untraced) Polyakov loop $\Omega(x)=
T\exp\left(-\int_{x_0}^{x_0+\beta}A_0(t,\bfx)dt\right)$
\cite{Garcia-Recio:2000gt}.

For our present purposes it will be sufficient to retain in
(\ref{eq:11}) only those terms with at most one spatial covariant
derivative. Upon substitution of these expressions in (\ref{eq:1}),
and taking the Dirac trace keeping only the abnormal parity terms,
produces
\begin{equation}
J_i^-(x)= 2 i m \epsilon_{ij}
\frac{1}{\beta}\sum_{p_0}\int\frac{d^2\bfp}{(2\pi)^2}
\langle x| \frac{1}{\Delta}E_j \frac{1}{\Delta}|0\rangle +  O(D_i^3) \,,
\label{eq:10}
\end{equation}
where $\Delta= m^2+\bfp^2+(p_0-iD_0)^2$. It is noteworthy that the
result is directly ultraviolet finite and also consistent, i.e. a true
variation
\footnote{Consistency requires $\int
d^3x\,\tr(\delta_1A_i\delta_2J_i)$ to be symmetric under the exchange
of the labels 1 and 2.}.

The current (\ref{eq:10}) picks up contributions from all terms with
two spatial indices in the effective action, however, most of those
terms are perfectly regular, in the sense that they are ultraviolet
finite, one-valued, strictly gauge invariant and anomaly free,
therefore we can simplify the calculation by isolating only the
anomalous terms, namely
\begin{equation}
J_i^{\text{an}}=  i m \epsilon_{ij}
\frac{1}{\beta}\sum_{p_0}\int\frac{d^2\bfp}{(2\pi)^2}
\langle x|\left\{ \frac{1}{\Delta^2},E_j\right\}|0\rangle \,.
\label{eq:2}
\end{equation}
This ``anomalous'' current is still consistent. The difference between
the two currents, $J_i^--J_i^{\text{an}}$, involves $[D_0,E_i]$ and
derives from (fully calculable) regular terms in the effective action
which are quadratic in $E_i$ \cite{Salcedo:2002prep}. On the contrary
the anomalous current does not contain such a commutator and so it
cannot derive from one of those terms. After carrying out the momentum
integration and the sum over frequencies, the anomalous current takes
the simple explicit form
\begin{eqnarray}
J_i^{\text{an}} &=& \epsilon_{ij}
\langle x|\left\{ \varphi(D_0),E_j\right\}|0\rangle \,,
\label{eq:3}
\end{eqnarray}
where we have introduced the function
\begin{equation}
\varphi(a)=\frac{i}{8\pi}
\frac{\sinh(\beta m)}{\cosh(\beta m)+\cosh(\beta a)} \,.
\end{equation}
This current preserves parity (as defined above), and can be written
in a manifestly gauge invariant form by using the identity $e^{-\beta
D_0}=\Omega(x)$ \cite{Garcia-Recio:2000gt}:
\begin{eqnarray}
J_i^{\text{an}} &=& \frac{i}{4\pi}\epsilon_{ij}
\left\{ 
\frac{\sinh(\beta m)}{2\cosh(\beta m)+\Omega+\Omega^{-1}}
,E_j\right\} \,.
\label{eq:3a}
\end{eqnarray}
Note that this current gives the static limit of the gauge field
self-energy instead of the long-wave limit \cite{Kao:1993fx}. The
latter limit follows from restoring the regular terms stripped from
the anomalous piece. Our criterion for selecting the anomalous piece
is that the remainder is strictly gauge invariant, however, regarding
the self-energy after rotation to real time, the anomalous piece is
analytical at zero momentum and energy and the non-analytical behavior
is contained in the remainder \cite{Salcedo:2002prep}.

At this step we can already verify that the zero temperature limit
comes out correctly. Indeed, for large $\beta$
\begin{equation}
J_i^{\text{an}} = \frac{i}{4\pi}\epsilon(m) \epsilon_{ij} E_j \,,\quad
(T=0)\,
\label{eq:0}
\end{equation}
(where $\epsilon(m)$ stands for the sign of $m$). This current derives from
\begin{equation}
W_{\text{an}} =  -\frac{1}{2}\epsilon(m) W_{\text{CS}} \,,\quad (T=0),
\label{eq:0a}
\end{equation}
where $W_{\text{CS}}$ is the Chern-Simons action
\begin{equation}
W_{\text{CS}}= \frac{i}{4\pi}\int
d^3x\epsilon_{ij}\tr(A_0F_{ij} - A_i\partial_0A_j) \,.
\end{equation}
Eq. (\ref{eq:0a}) is the standard induced Chern-Simons term at zero
temperature \cite{Pisarski:1987gq,GamboaSaravi:1996aq,Babu:1987rs}.

\section{The effective action}
Using the techniques developed in
\cite{Salcedo:1998sv,Garcia-Recio:2000gt} it is possible to
reconstruct the anomalous piece of the effective action from its
current while preserving manifest gauge invariance at every step of
the calculation. This method will be presented elsewhere
\cite{Salcedo:2002prep}. A more direct approach, which we will follow
here, is to proceed by fixing the gauge \footnote{The gauge conditions
(\ref{eq:5a}) and (\ref{eq:5b}) depend only on $A_0$ and so they do
not interfere with an arbitrary local variation of $A_i(x)$, as
required to obtain the current and reconstruct the effective action
from it.}. We choose the gauge so that $A_0(x)$ is stationary
\begin{equation}
\partial_0 A_0(x)= 0\,.
\label{eq:5a}
\end{equation}
Such a gauge always exists (globally) \cite{Salcedo:1998sv}. In this
gauge, (\ref{eq:3a}) becomes
\begin{equation}
J_i^{\text{an}}(x) = \epsilon_{ij} \left\{
\varphi(A_0(\bfx)),E_j(x)\right\} \,.
\label{eq:4}
\end{equation}
The final step is to recover the effective action associated to this
current. To this end we split the current into two parts
$J_i^{(1)}+J_i^{(2)}$, corresponding to the decomposition of the
electric field into $E_i=[D_0,A_i]-\partial_i A_0$. Noting that in our
gauge $A_0$ and $D_0$ commute, it is readily verified that the first
piece derives from
\begin{equation}
W^{(1)}= \int d^3x \epsilon_{ij} \tr\left( \varphi(A_0) \,
A_i[D_0,A_j] \right) \,.
\label{eq:7a}
\end{equation}
Thus the remainder must satisfy
\begin{equation}
\delta W^{(2)}= 
-\int d^3x \epsilon_{ij} \tr\left(\delta A_i
\left\{ \varphi(A_0) , \partial_j A_0 \right\}
\right)
\label{eq:6}
\end{equation}
To proceed, we further restrict the gauge (by means of a suitable
subsequent stationary gauge transformation) so that $A_0(\bfx)$ is
everywhere a diagonal matrix, i.e
\begin{equation}
[A_0(\bfx),\partial_i A_0(\bfx)]= 0\,.
\label{eq:5b}
\end{equation}
For subsequent reference we note that in this gauge
\begin{equation}
[D_0,[D_0,A_i]]= [D_0,E_i]\,.
\label{eq:21}
\end{equation}

With this choice of gauge, (\ref{eq:6}) has the simple solution
\begin{equation}
W^{(2)}=  -\int d^3x \epsilon_{ij} \tr\left(
\Phi(A_0)\partial_i A_j  \right)
\label{eq:7}
\end{equation}
provided that $\Phi^\prime(a)= 2\varphi(a)$ or, in our case,
\begin{eqnarray}
\Phi(a)=\frac{i}{2\pi\beta}
\tanh^{-1}\left(\tanh(\beta m/2 ) \tanh(\beta a/2) 
\right)  \,.
\end{eqnarray}
Using the convenient notation of differential forms (for the spatial
indices only), the final result obtained by adding up (\ref{eq:7a})
and (\ref{eq:7}), takes the form
\begin{equation}
W_{\text{an}}= \int dx_0 \,\tr\left( 
\Phi(A_0)(A^2-B)
+\varphi(A_0) \,A[D_0,A] 
\right) \,
\label{eq:8}
\end{equation}
(where $A=A_idx_i$ and $B=dA+A^2= \frac{1}{2}F_{ij}dx_idx_j$ is the
magnetic field). This action reproduces the spatial components of the
current. Since any further contributions to the effective action
cannot contain $A_i$, they would be of the form $ \int dx_0
\,\tr\left( f(A_0)dA_0dA_0 \right) $ and this vanishes identically in
our gauge.

A subtle, but important, point is that, depending on the space
manifold and the gauge group, a given configuration $A_0(\bfx)$ may
not be globally diagonalizable (i.e. with continuity at all points).
Nevertheless, the ``diagonal'' gauge (\ref{eq:5b}) always exists
locally, that is, for each of the patches covering the space manifold.
Without loss of generality we can assume that the support of the local
variation in (\ref{eq:6}) takes place inside one of the patches, thus
integration by parts in $\bfx$ is allowed and the correct current is
obtained from (\ref{eq:7}). Because the current (\ref{eq:4}) is patch
independent, this suggests that the integrand in (\ref{eq:8}) is also
patch independent. As shown in the next section, this is actually
correct, confirming that the functional $W_{\text{an}}$ is well
defined. For simplicity, we will momentarily assume $A_0(\bfx)$ to be
diagonal in a global gauge.

Eqs.~(\ref{eq:0}) and (\ref{eq:0a}) show that the calculation is
consistent with the known zero temperature limit, and this can also be
verified directly from $W_{\text{an}}$. Next we can study the behavior
of this functional under gauge transformations.  We have to
distinguish between the ``gauge fixing'' transformation needed to
bring the original configuration to the gauge where $A_0$ is
stationary and diagonal, on the one hand, and the allowed gauge
transformations which preserve the gauge conditions (\ref{eq:5a}) and
(\ref{eq:5b}), on the other. Of course, gauge invariance of the
partition function under the gauge fixing transformation cannot be
studied using our functional. Such gauge invariance follows from
general arguments, e.g. by using the $\zeta$-function version of the
effective action \cite{Deser:1997nv}. The gauge conditions
(\ref{eq:5a}) and (\ref{eq:5b}) are preserved by two kinds of gauge
transformations, namely, stationary and discrete transformations
\cite{Salcedo:1998sv}. The first class is that of transformations
which are stationary and diagonal, and it is easily verified that
$W_{\text{an}}$ is invariant under such transformations. The second
class refers to those transformations of the form
$U=\exp(x_0\Lambda)$, where $\Lambda$ is a constant diagonal matrix
with entries $\lambda_j=2\pi in_j/\beta$ (with $n_j$ integer).  Under
discrete transformations $A_i$ transforms covariantly (i.e.,
homogeneously) whereas $A_0\to A_0+\Lambda$. The current (\ref{eq:4})
transforms covariantly (due to periodicity of $\varphi(a)$) thus
$W_{\text{an}}$ can only change by constant additive terms, that is,
terms that do not give a contribution to the transformed current. This
can be verified directly from the functional as follows. First, at a
formal level the function $\Phi(a)$ is periodic and $W_{\text{an}}$ is
invariant. More correctly, on its Riemann surface $\Phi(A_0+\Lambda)=
\Phi(A_0)+i\epsilon(m)\Lambda/4\pi$ and this induces a variation
$-i\epsilon(m)/4\pi \int dx_0 \,\tr ( \Lambda dA)$ on
$W_{\text{an}}$. This variation is precisely the same as that computed
for the zero temperature functional in (\ref{eq:0a}) under the same
gauge transformation. This implies that this variation is of the form
$\pm i\pi kn$, where $k$ is an integer depending on the gauge group
and the gauge field configuration and $n$ is the winding number of the
gauge transformation \footnote{As introduced here, discrete
transformations are always topologically small for simply connected
groups, however, this is not necessarily the case when the diagonal
gauge is not global (see section \ref{sec:topo}).}. It can be noted
that a perturbative expansion in $W_{\text{an}}$ or $J_i^{\text{an}}$
would break periodicity of the functions $\varphi$ and $\Phi$ and, in
general, this spoils gauge invariance under discrete transformations.

If there are large transformations and $k$ is odd, the partition
function changes by a factor $(-1)^n$ and gauge invariance must be
enforced by adding a polynomial term $\pm\frac{1}{2}W_{\text{CS}}$
\cite{Redlich:1984kn}. Such term is present directly in the effective
action when the $\zeta$-function regularization is used
\cite{GamboaSaravi:1996aq,Salcedo:1998sv}. This term is even as a
function of $m$ and so it introduces an anomalous breaking of
parity. Note that this new term also adds a parity breaking
contribution to the current; although the current (\ref{eq:3}) is
gauge and parity invariant it does not derive from a gauge invariant
partition function. It also noteworthy that for massless fermions the
factor $(-1)^n$ is also present \cite{Redlich:1984kn} but comes from
the normal parity component of the effective action, since
$W_{\text{an}}$ vanishes in this case.

In \cite{Deser:1997nv,Fosco:1997ei} it is shown that the full
(abnormal parity) result is $-\int dx_0 \,\tr\left(\Phi(A_0)B \right)$
provided that the configuration is stationary, Abelian and $A_0$ is
$\bfx$-independent. The same formula is found in \cite{Brandt:2001ni}
in the non-Abelian stationary case when the electric field vanishes.
This is consistent with our $W_{\text{an}}$ noting that in both cases
the $A_i$ are stationary and commute everywhere with $A_0$, and so the
explicit $A$ cancel in (\ref{eq:8}).

\section{Topological issues}
\label{sec:topo}
Let us now analyze the general case in which the diagonal gauge is not
assumed to exist globally. In this case it will be convenient to work
in the global gauge where $A_0$ is stationary but not necessarily
diagonal, and rewrite $W_{\text{an}}$ in that gauge. This yields
\begin{equation}
W_{\text{an}}= \int dx_0 \,\tr\left( 
\Phi(A_0)(\A^2-B)
+\varphi(A_0) \,\A[D_0,\A] 
\right) \,
\label{eq:8a}
\end{equation}
In this formula $\A$ denotes the vector field $A$ of the diagonal
gauge rotated covariantly back to the stationary gauge. That is, if
$U$ denotes the stationary gauge transformation which brings $A_0$ to
a (local) diagonal gauge,
\begin{equation}
\A= A+dUU^{-1}\,.
\label{eq:19}
\end{equation}
A crucial property of (\ref{eq:8a}) is that it remains unchanged under
the replacement
\begin{equation}
\A \to \A^\prime=\A+C\,,
\label{eq:20}
\end{equation}
provided that the field $C(x)$ satisfies
\begin{equation}
\partial_0 C= [A_0,C]=0 \,.
\label{eq:20a}
\end{equation}
This is readily verified noting that $C$ is a 1-form and $A_0$, $D_0$
and $C$ are all commuting quantities.

By construction $\A$ transforms covariantly (under transformations of
the global stationary gauge and fixed local diagonal gauge), but is
not globally defined since the gauge transformation $U$ only exists
locally. Let $M^{(k)}$ denote a set of local charts covering the space
manifold. In each chart we can take a diagonalizing gauge
transformation $U^{(k)}$ and this defines a corresponding ``covariant
vector field'' $\A^{(k)}$. What have to be verified is that 2-form in
(\ref{eq:8a}) takes the same value in any of the charts. 

To show this let us note that the relation (\ref{eq:21}), which was
written in the diagonal gauge, in the stationary gauge becomes
\begin{equation}
[D_0,[D_0,\A^{(k)}]]= [D_0,E]\,.
\label{eq:21a}
\end{equation}
Therefore the field $C^{(k,\ell)}= \A^{(k)}-\A^{(\ell)}$ (which is
stationary due to (\ref{eq:19})) satisfies
\begin{equation}
[A_0,[A_0,C^{(k,\ell)}]]= 0\,.
\end{equation}
This already implies that $C^{(k,\ell)}$ commutes with $A_0$ and
satisfies (\ref{eq:20a}). As a consequence any two $\A^{(k)}$ at the
same point are related as in (\ref{eq:20}) and $W_{\text{an}}$ is
patch independent.

It is noteworthy that the field $A_0^\prime(\bfx)$, denoting $A_0$ in
a diagonal gauge, is (or can be taken to be) equal in all
patches. Indeed, the local eigenvalues of $A_0(\bfx)$,
$a_\lambda(\bfx)$, can be labeled at each point so that they are
continuous functions on the space manifold. Thus, without loss of
generality we can choose the local diagonal gauges in such a way that
the eigenvalues are ordered in all patches in the same way and so the
$A_0^\prime(\bfx)$, which are diagonal matrices containing the
eigenvalues, will be the same matrix in all patches. However, although
$A_0^\prime(\bfx)$ is global, it needs not correspond to a global
gauge; in general, the eigenvectors corresponding to $a_\lambda(\bfx)$
cannot be chosen in a continuous way on the space manifold. Assuming
the non degenerated case for simplicity, the choice of eigenvectors in
each patch will differ by a phase (and perhaps normalization,
depending on the gauge group). These phases are contained in the
diagonal transition matrices ${U^{(\ell)}}^{-1}U^{(k)}$, which contain
the non trivial topology of $A_0(\bfx)$ (under gauge transformations).

When we bring a completely general gauge field configuration to an
$A_0$-stationary gauge the result is not unique. The different
possible stationary gauges so obtained are related either by a
stationary gauge transformation or by a discrete transformation
\cite{Salcedo:1998sv}. The discrete transformations are of the form
$U=\exp(x_0\Lambda)$, where $\Lambda(\bfx)$ is stationary, commutes
with $A_0(\bfx)$ and has eigenvalues $\lambda_j=2\pi in_j/\beta$ (with
$n_j$ integer). The discussion of how $W_{\text{an}}$ changes under
these transformations is essentially the same as that given before for
the case of a global diagonal gauge. It can be noted, however, that
the discrete transformations need not be topologically small if the
diagonal gauge is not global (an explicit SU(2) example is shown in
\cite{Pisarski:1987gq}, also discussed in \cite{Salcedo:1998sv}.)

\section{Conclusions}
In summary, we have shown that the method of expanding in the number
of spatial covariant derivatives is suitable to make computations at
finite temperature, and in particular it is able to retain all
subtleties usually tied to the effective action, such as topological
pieces, many-valuation and anomalies. At the practical level, the
method have been shown to be quite efficient, yielding a simple
explicit form for the induced CS term at finite temperature, without
assuming special constraints on the gauge field configuration.

\begin{acknowledgments}
This work is supported in part by funds provided by the Spanish DGICYT
grant no.  PB98-1367 and Junta de Andaluc\'{\i}a grant no. FQM-225.
\end{acknowledgments}


\end{document}